\documentclass[12pt]{article}

\usepackage{graphicx}

\begin{document}

\title{Approach to the Rotation Driven Vibrational to
Axially Rotational Shape Phase Transition Along the Yrast Line of
a Nucleus}

\author{{Yu-xin Liu$^{1,2,3,4}$, Liang-zhu Mu$^{1}$,
and Haiqing Wei$^{5,6}$ }\\[3mm]
\normalsize{$^1$ Department of Physics, Peking University, Beijing 100871,
China}\\[-1mm]
\normalsize {$^{2}$ Key Laboratory of Heavy Ion Physics,
Ministry of Education, Beijing 100871, China } \\[-1mm]
\normalsize{$^3$ Institute of Theoretical Physics, Academia
Sinica, Beijing 100080, China} \\[-1mm]
\normalsize{$^4$ Center of Theoretical Nuclear Physics, National
Laboratory of}\\[-1mm] \normalsize{ Heavy Ion Accelerator, Lanzhou
730000, China}  \\[-1mm]
\normalsize{$^5$ Allambda Technologies, Inc., San Jose, CA 95134,
USA}\\[-1mm]
\normalsize{$^6$ School of Information Science and Engineering,
Lanzhou University, Lanzhou 730000, China} }


\maketitle

\begin{abstract}
By analyzing the potential energy surface, the shape phase diagram
and the energy spectrum of the nucleus in U(5) symmetry in the
IBM, we propose that the U(5) symmetry with parameters $(A+B) < 0$
may be a model to describe the rotation driven vibrational to
axially rotational phase transition along the yrast line. With
such a model, we have described successfully the observed rotation
driven shape phase transition along the yrast line of individual
nucleus and proposed some other empirical evidences.
\end{abstract}





\newpage

\parindent=20pt

It has been well known that shape phase transition is one of the
most significant topics in nuclear structure research. Many
evidences of nuclear shape phase transition have been observed.
For instance, in several isotopes, there exists shape phase
transition from vibration to axial rotation or $\gamma$-unstable
rotation with respect to the variation of neutron
number\cite{IA87}, and a triple point may
appear\cite{Jolie02,Warn02}. Even in one mode of collective motion
there may involve different characteristics, for example, along
the yrast line there exists transition between rotations holding
different relations between angular momentum and rotational
frequency (referred to as band crossing and exhibiting a
backbending)\cite{SS72}. Very recently it was found that there
involves rotation driven vibrational to axially rotational shape
phase transition along the yrast line of individual
nucleus\cite{Regan03}. On the theoretical side, the interacting
boson model (IBM) has been shown to be successful in describing
the shape phase transition along a series of
isotopes\cite{IA87,Cejn03,IZ04}. And analytic solutions for the
critical points of the phase transitions have been
found\cite{Iac00,Iac01,LG03}. The cranked shell model
(CSM)\cite{BRM86} has been known to be able to describe the band
crossing very well. However, a theoretical approach to describe
the rotation driven shape phase transition from vibration to axial
rotation along the yrast line in individual nucleus has not yet
been established, even though several attempts have been made (see
for example Ref.\cite{Cejn04}). By analyzing the potential energy
surface and the energy spectrum of the U(5) symmetry in the IBM,
we will show that the U(5) symmetry with a special choice of
parameters can be a model to describe the rotation driven
vibrational to axially rotational shape phase transition along the
yrast line of individual nucleus.

In the original version of IBM (IBM1), the collective motion of a
nucleus is described by the coherent state of $s$- and $d$-bosons,
which hold angular momentum 0, 2, respectively. The corresponding
symmetry group is U(6), and possesses three dynamical symmetry
limits, namely, U(5), O(6) and SU(3). With one- and two-body
interactions among the bosons being taken into account, the
Hamiltonian of the U(5) symmetry can be written as\cite{IA87},
$$\displaylines{\hspace*{1cm}
\hat{H}_{U(5)}=E_0 + \varepsilon_{d}  C_{1U(5)} + A C_{2U(5)} + B
C_{2O(5)}+ C C_{2O(3)} \, , \hfill{(1)} \cr }
$$
where $C_{kG}$ is the $k$-rank Casimir operator of group $G$, and
the parameters satisfy $C\ll | B | \ll \vert A \vert \ll
\varepsilon_d$. The IBM may be linked to the collective model of
Bohr and Mottelson by implementing the coherent state
formalism\cite{IA87,GK80,DSI80,IC81}. In IBM1, the intrinsic
coherent state of a nucleus with $N$ bosons is given by
\setcounter{equation}{1}
$$\displaylines{\hspace*{3cm} \left|
{N;\beta ,\gamma } \right\rangle  = \left[ {s^{\dag} +
\sum\limits_\mu  {\alpha _\mu  d_\mu ^ {\dag}  } } \right]^N
\left| 0 \right\rangle \, , \hfill{(2)} \cr }$$
 with
$$ \alpha_0=\beta \cos \gamma, \quad \alpha_{\pm 1}=0,
\quad \alpha_{\pm 2}=\frac{1}{\sqrt 2} \beta \sin \gamma \, . $$
 It is evident that all components of $d$-bosons can be described
by the two parameters $\beta$ and $\gamma$, which have been shown
proportional to the deformation parameters in the collective model
\cite{IA87}. For instance, the relations $\hat{\beta}_{2} \approx
0.15 \beta$ and $\hat{\gamma} = \gamma$ hold for rare earth
nuclei, where $\hat{\beta}_{2}$ and $\hat{\gamma}$ are the
quadrupole deformation parameters in collective model. We then
refer to the parameters $\beta$ and $\gamma$ as deformation
parameters in the follows.

It is remarkable that the coherent state of Eq.~(2) consists of
all possible components of $d$ bosons. As a consequence, it is not
an eigenstate of angular momentum, but a superposition of states
with all possible values of angular momentum. To investigate the
shape phase structure and its transition among the states in
ground-state band (or yrast band), we implement the technique of
angular momentum projection \cite{KM88,Dobes90,HS95} involving
projection operator
$$\displaylines{\hspace*{3cm}
P^L_{MK}=\frac{2L+1}{8 \pi ^2} \int
{D^L_{MK}}^*(\Omega)R(\Omega)d\Omega \, , \hfill{(3)} \cr }
$$
 with $M=K=0$. Then the energy functional ({\it i.e.} potential
energy surface) of the state with angular momentum $L$ in the
ground-state band can be given as
$$\displaylines{\hspace*{2cm} E_{gsb}(N,L,\beta ,
\gamma)  = \frac{{\left\langle {N;\beta ,\gamma } \right| \hat{H}
P_{00}^L \left| {N;\beta ,\gamma } \right\rangle }}{{\left\langle
{N;\beta ,\gamma } \right|P_{00}^L \left| {N;\beta ,\gamma }
\right\rangle }} \, . \hfill{(4)} \cr }$$
 With the rotation operator written explicitly, Eq.~(4) becomes
$$\displaylines{\hspace*{2cm} E _{gsb}(N,L,\beta , \gamma) =
\frac{{\int^{\pi}_0 {d\beta '\sin \beta 'd_{00}^L (\beta
')\left\langle {N;\beta ,\gamma } \right| \hat{H} e^{ - i\beta
'J_y } \left| {N;\beta ,\gamma } \right\rangle } }}{{\int^{\pi}_0
{d\beta '\sin \beta 'd_{00}^L (\beta ')\left\langle {N;\beta
,\gamma } \right|e^{ - i\beta 'J_y } \left| {N;\beta ,\gamma }
\right\rangle } }} \, , \hfill{(5)} \cr }$$ where $d^L_{mm'}(\beta
')$ is the reduced rotation matrix.

After some derivations, and considering that the deformation
parameter $\beta$ should be rather small for the vibrational to
axially rotational ($\gamma =0$) shape phase transition of current
interest, we can approximate the potential energy surface of
Eq.~(5) up to $\beta^6$ as
$$\displaylines{\hspace*{25mm}  E _{gsb}(N,L,\beta ) =
A_0+ \frac{1}{2} \alpha (L - L_0) \beta^2 + \frac{1}{4} A_4
\beta^4 + \frac{1}{6} A_6 \beta^6  \, ,  \hfill{(6)} \cr }$$
 with
 $$ A_0(N,L,\varepsilon_{d},A,B,C) = \frac{1}{4}
\left[ 2 \varepsilon_{d} + 8 A + 6B + 4C + (A + B + 4C) L \right]
L \, ,
$$
 $$ \alpha = \frac{L(2N-L) (A + B) } {2(3+2L)}\, ,$$
$$ L _0 =  \frac{\varepsilon_{d} + 5A + 4B}{-(A + B )} \, .$$
$A_4$ and $A_6$ are also functions of the interaction parameters
$\varepsilon_{d}$, $A$, $B$, $C$ and the state indices $N$, $L$.

It follows from the spectrum generating principle that $L \in [0,
2N]$, $\varepsilon _{d} > 0$, while $A$, $B$ and $C$ may be either
positive or negative. If $(A+B)>0$, then $\alpha > 0$ and $L_{0}
<0$, so that $\alpha (L - L_0 )$ remains positive. Otherwise, if
$(A + B) < 0$ ( the dynamical symmetry condition guarantees
$(\varepsilon_{d} + 5 A + 4B) > 0$ ), then $\alpha<0$ and $L_0>0$,
so that $\alpha (L - L_0)$ may change from positive to negative as
the angular momentum $L$ increases from under to over $L_0$.
Because of the complexity of the $A_{4}$ and $A_{6}$ in terms of
the parameters $N$, $L$, $\varepsilon_{d}$, $A$, $B$  and $C$, it
is difficult to discuss the variation characteristics
analytically. Numerical calculation indicates that, when the
parameters in the Hamiltonian are taken as $\varepsilon_{d}>0$,
$(A + B) <0 $ and $C >0$,  the parameters $A_{4}$, $A_{6}$ in
Eq.~(6) can be negative, positive, respectively. It becomes then
evident that the energy functional of Eq.~(6) closely resembles
the Landau free-energy \cite{LL01,KKbook} in the theory of
thermodynamic phase transition, with $L$ and $\beta$ playing the
roles of control parameter and order parameter, respectively.

A more quantitative analysis is now in order for the potential
energy surface of Eq.~(6). If the parameters in the Hamiltonian
(1) are chosen as  $(A+B)>0$, we have $\frac{\partial^2
E_{gsb}(N,L,\beta)}{\partial\beta^2} \vert_{\beta=0}=\alpha (L -
L_0)>0$. Then the potential energy is minimized only at $\beta=0$.
The nucleus with such a potential energy functional would always
stay in a vibrational shape phase. If $(A+B)<0$, then $\alpha<0$
and $L_0>0$, so $\frac{\partial ^2 E_{gsb}(N,L,\beta )}{\partial
\beta ^2} \vert _{\beta =0} > 0 $ for $L<L_0$, and $\frac{\partial
^2 E_{gsb}(N,L,\beta )}{\partial \beta ^2} \vert _{\beta =0} < 0 $
for $L > L_0$. The point $\beta=0$ evolves from a local minimum to
a local maximum of the potential energy surface when $L$ rises
from below to above $L_0$. Furthermore, there exists a critical
angular momentum $L_c = L_0 + \frac{3 A_4 ^2}{16 A_{6} \alpha }$,
with which the energy surface acquires two and equal minima at
$\beta =0$ and $\beta=\sqrt{-3A_4/4A_6}$, respectively. There is
another critical value $L_{max}$ for the angular momentum, beyond
which a nucleus becomes unstable as the energy functional is no
longer lower-bounded. It follows from Eq.~(6) that $L_{max}=2N$,
the highest angular momentum possible, when $A_6>0$, whereas
$L_{max}=L_0 + \frac{A_4^2}{4A_6\alpha}$ when $A_6<0$. In any
case, we have $L_{max}>L_0$. For the states with $L\in
(L_{c},L_{max}]$, the energy functional is maximized at $\beta=0$
and minimized at some $\beta\ne 0$. As already mentioned, the
energy functional has no minimum for the states with angular
momentum $L\in( L_{max},2 N]$. The $\beta$-dependence of the
energy surface for some typical values of angular momentum $L$ is
illustrated in Fig.~1, where $L_0$ and $L_c$ are calculated as 8
and 6 respectively. It becomes apparent that a nucleus has a
stable vibrational shape phase with states around $\beta=0$, if
the Hamiltonian parameter $(A+B)<0$ and the angular momentum
$L<L_c$. In the regime of $L\in (L_{c}, L_{max}]$, the only stable
shape phase is rotational with states around the energy minimum at
some $\beta \ne 0$. For $L= L_c$, the energy surface has two
degenerate but distinct minima, one of which is localized at
$\beta=0$, the other is localized at some $\beta\ne 0$. Under such
a condition, the system may undergo a transition from one energy
minimum to the other, and when that happens, the symmetry of the
system is spontaneously broken. Such transition is of the first
order, according to the standard theory of phase transition. The
same theory would also predict the coexistence of vibrational and
rotational shapes with $L= L_c$ as a precursor of a shape phase
transition\cite{SRJL03,Heyde04}. When the angular momentum
$L>L_{max}$, a nucleus may not be able to maintain a stable
structure. Fig.~2 depicts the shape phase diagram of a nucleus in
terms of the angular momentum $L$ and the deformation parameter
$\beta$, where the same interaction parameters as in Fig.~1 are
used, $L_0$ and $L_c$ still are 8 and 6, respectively.

The above discussion has shown explicitly that the potential
energy functional derived from the U(5) symmetry with $(A+B)<0$
has the similar mathematical form of the Landau
free-energy\cite{LL01,KKbook}, which puts a U(5)-symmetric nuclear
system in the standard theoretical framework of first-order phase
transition, so to correctly predict and well describe the
vibrational to axially rotational nuclear shape phase transition
in the ground-state band (or along the yrast line).

On the phenomenological side, the energy spectrum of a nucleus in
U(5) symmetry can be given as
$$\displaylines{\hspace*{1cm}
E_{U(5)}=E_0 + \varepsilon _{d} n_d + A \, n_d(n_{d}+4) + B\,
\tau(\tau+3)+ C \, L(L+1) \, . \hfill{(7)} \cr }
$$
where $n_d$, $\tau$, and $L$ are the irreducible representations
(IRREPs) of the group U(5), O(5) and O(3), respectively.

From the spectrum generating process one can recognize that, if
$A>0$ and $B>0$, the states with $\tau = n_{d}$, $L = 2 n_{d}$
form the ground state band and simultaneously the yrast band, and
the energy of the state in the ground state band can be given as
$$E_{gsb}(n_d) = E_{0} + (A + B ) {n_{d}}^2
+ (\varepsilon_{d} + 4A + 3B ) n_{d}  + C L (L +1) \,  .   $$
 The energy spectrum appears as the anharmonic vibrational one with
increasing frequency $\hbar\omega= (\varepsilon _{d}+ 4A + 3B ) +
(A + B ) n_{d} $ (if the anharmonic effect induced by the rotation
is taken into account, the vibrational frequency increases with
$n_{d}$ in the relation  $\hbar\omega= (\varepsilon _{d}+ 4A + 3B
+ 2C) + (A + B + 4C) n_{d} $). {\it However}, if $(A+B)<0$, the
$E_{gsb}(n_d)$ ($E_{gsb}(L)$) is an upper-convex parabola against
the $n_d$ ($L$) and appears as the anharmonic vibrational one with
decreasing frequency. From these anharmonic vibrational
characteristics, one can recognize that, when $(A+B) < 0$, there
exists a d-boson number $n^{(c)}_{d}$ and an angular momentum
$$\displaylines{\hspace*{1cm}
L_{c}=2n_{d}^{(c)} = - \frac{2(\varepsilon_{d} + 4 A + 3 B )}{A +
B } - 2N_{0} \, , \hfill{(8)} \cr }
$$
where $N_0=N$ with $N$ being the total boson number. As the
angular momentum $L \geq L_c$, the yrast states are no longer the
anharmonic vibrational ones mentioned above, but the
quasi-rotational ones with $ n_{d}= N_0$. Recalling the analysis
in the coherent state formalism with angular momentum projection,
one can conclude that the U(5) symmetry with parameters $(A+B)<0$
can describe the vibrational to axially rotational shape phase
transition along the yrast line. The $L_c$ given in Eq.~(8) is the
critical angular momentum. For each yrast state with angular
momentum $L < L_{c}$, its energy can be given as
$$\displaylines{\hspace*{1cm}
E(L < L_c)=E_0 + \frac{ A + B + 4 C}{4} {L}^{2} +
\frac{\varepsilon_{d} + 4 A + 3 B + 2 C} {2} L \, . \hfill{(9)} \,
\cr }
$$
Whereas for the one with angular momentum $L \ge L_{c}$, its
energy should be expressed as
$$\displaylines{\hspace*{1cm}
E(L \ge L_{c} ) = E_{0} ^{\prime } + C L ( L +1)  \, ,
\hfill{(10)} \cr }
$$
and can be displayed as a part of an upper-concave parabola. For
instance, for a system with $N=15$ and parameters $\varepsilon
_{d}=0.80$~MeV, $A = -0.025$~MeV,  $ B = -0.01$~MeV and $ C =
0.004$~MeV, with Eq.~(8) we can fix the critical angular momentum
$L_c=8 \hbar $. The energy of the yrast states against the angular
momentum $L$ can be illustrated in the left panel of Fig.~3. It
has been known that the energy of E2 transition $\gamma$-ray over
spin (E-GOS)  $ R = \frac{E_{\gamma}(L \rightarrow L-2)}{L} $ can
be taken as a quite good signature to manifest the vibrational to
axially rotational phase transition along the yrast
line\cite{Regan03}. As an auxiliary evidence, we show also the
E-GOS of the yrast states with above parameters in the right panel
of Fig.~3. The figure indicates apparently that the yrast states
involve a vibrational to axially rotational phase transition and
the angular momentum $L_c$ is definitely the critical point for
the phase transition to take place. Such a transition is quite
similar to that between the states with $n_p=0$ and
$n_p=N$\cite{BF02} in the vibron model\cite{I78} with random
interactions.

Reviewing above analysis, one knows that, in the anharmonic
vibrational model, there involves competition between vibration
and rotation. In the case of that the interaction parameters are
taken as $(A + B) < 0$, the vibrational frequency decreases and
the rotational effect increases if the angular momentum (or $d$
boson number) increases. As the angular momentum $L$ (or $n_{d}$)
reaches the critical value, the vibration disappears, so that only
the rotational effect governs the property of state.
Simultaneously, a sudden increase happens in the $d$ boson number.
Then the structure of the wavefunction changes from the
vibrational one to the rotational one, so that the vibrational to
the axially rotational shape phase transition takes place. In the
microscopic point of view, cranking random phase approximation
calculation has shown that the gradual decrease of vibration can
induce a backbending, i.e., a shape phase transition\cite{Naz04}.
On phenomenological level, the U(5) symmetry with parameter $A <
0$ has been used to describe the collective backbending of high
spin states successfully\cite{Long97}.

As an application of our presently proposed model, we analyze the
typical example $^{102}$Ru involving the rotation driven
vibrational to axially rotational shape phase
transition\cite{Regan03}. By fitting the experimental data of the
yrast band of $^{102}$Ru with Eqs.(9) and (10), we obtain the
energy spectrum and the E-GOS plots as shown in Fig.~4. To show
the sudden change of the wavefunction, we also list the
configuration $(n_{d}, \tau, L)$ of the sates in the yrast band in
Fig.~4 and the fitted parameters in Table 1. The figure and the
Table show evidently that the shape phase transition happens at
angular momentum $L=12$ and our model describes well such a shape
phase transition.

\begin{table}[ht]
\begin{center}
\caption{Fitted parameters of the bands $^{102}$Ru, $^{112}$Cd,
$^{114}$Cd, $^{114}$Te, $^{142}$Sm and $^{188}$Hg }
\begin{tabular}{|c|c|c|c|c|c|c|}
\hline\hline band & $\varepsilon_{d}$ (MeV) & $A$ (MeV) & $B$
(MeV) & $C$ (MeV) & N & $L_{c}$  \\                \hline
$^{102}$Ru & $0.5281$ & $-0.009376$ & $-0.008819$ & $0.01475$ &
$20$ & $12$  \\
$^{112}$Cd & $0.6581$ & $-0.007640$ & $-0.01327$ & $0.01440$ &
$25$ & $10$  \\
$^{114}$Cd & $0.5695$ & $-0.0003789$ & $-0.01586$ & $0.01393$ &
$29$ & $10$  \\
$^{114}$Te & $0.7747$ & $-0.01678$ & $-0.01423$ & $0.01269$ &
$17$ & $12$  \\
$^{142}$Sm & $1.101$ & $-0.06263$ & $-0.01165$ & $0.01713$ &
$7$ & $10$  \\
$^{188}$Hg & $0.5957$ & $-0.03057$ & $-0.002976$ & $0.009193$ &
$9$ & $12$  \\ \hline \hline
\end{tabular}
\end{center}
\end{table}

We have also analyzed the available experimental energy spectra of
the yrast bands of even-even nuclei with $ 30\le Z \le 100$ and
simulated the data with least-square fitting in our present model.
We find that, besides the yrast bands of nuclei $^{102}$Ru,
$^{112}$Cd and $^{114}$Cd identified by Regan and collaborators in
Ref.\cite{Regan03}, the yrast bands of nuclei $^{114}$Te,
$^{142}$Sm and $^{188}$Hg involve the vibrational to axially
rotational shape phase transition. The theoretical results of the
energy spectra and the E-GOS plots of these bands and the
comparison with experimental data are illustrated in Fig.~5. The
fitted parameters are listed in Table~1. From Fig.~5 and Table~1,
one can infer that our model can describe the rotation driven
shape phase transition successfully.

Looking over the fitted boson number listed in Table 1, one may
know that, for $^{142}$Sm and $^{188}$Hg, it agrees well with the
simple ansatz: boson number is half of the valence nucleons (or
holes). However, for the nuclei in $A \sim 110$ mass region, it
differs from half of the valence nucleons obviously. Recalling the
recent suggestion that the valence neutrons may be in the
$h_{11/2}$ orbital\cite{Regan03,Regan95}, we infer that the
microscopic configuration of the nuclei in $A \sim 110$ mass
region is so complicated that the boson number of them can not be
simply taken as half of the valence nucleons. It means that the
effective boson number may be important to the structure of these
nuclei. In fact, quite early IBM
calculations\cite{Sam82,Heyde82,Scholt83,Heyde85} had shown that,
to describe the spectroscopic property of these nuclei well,
effective boson number should be implemented. Then we believe that
the fitted boson number is reasonable, even though it is not
consistent with half of the valence nucleons for the $A\sim 110$
nuclei. Meanwhile it provides a clue that the structure of the
nuclei involving rotation driven vibrational to axially rotational
shape phase transition is very complicated and needs sophisticated
investigation.

In summary, by analyzing the potential energy surface of the
nucleus in U(5) symmetry in the coherent state  formalism with
angular momentum projection in the IBM, we give a phase diagram of
the vibration and the rotation in terms of the angular momentum
and the deformation parameter in this letter. We have then
proposed that the U(5) symmetry with parameters $(A+B) < 0$ may be
a model to describe the rotation driven vibrational to axially
rotational shape phase transition along the yrast line. With such
a model, we have described successfully the vibrational to axially
rotational phase transition along the yrast line in $A \sim 110$
mass region identified by Regan and collaborators\cite{Regan03}.
By analyzing the available experimental spectra of even-even
nuclei with $ 30 \le Z \le 100$, we show that, besides the ones in
$A \sim 110$ mass region, the yrast band of nuclei $^{114}$Te,
$^{142}$Sm and $^{188}$Hg may also be the empirical evidences
involving the rotation driven vibrational to axially rotational
shape phase transition.

\bigskip

This work is supported by the National Natural Science Foundation
of China under the contract Nos. 10425521, 10075002, and 10135030,
the Major State Basic Research Development Program under contract
No. G2000077400 and the Research Fund for the Doctoral Program of
Higher Education of under Grant No. 20040001010. One of the
authors (Y.X. Liu) thanks the support by the Foundation for
University Key Teacher by the Ministry of Education, China, too.


\newpage

\parindent=0pt


\begin{figure}
\begin{center}
\includegraphics[scale=1.20,angle=0]{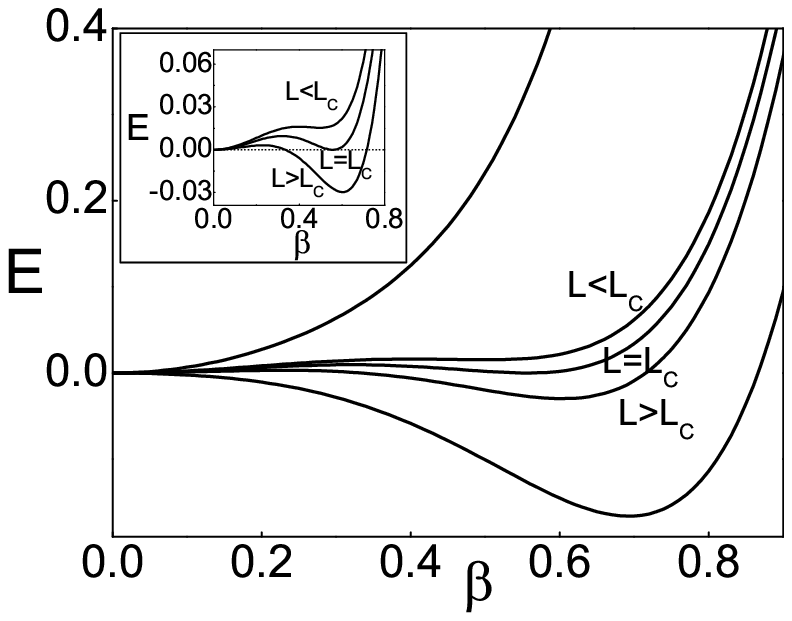}
\caption{The energy surface (Landau free energy) of a nucleus
against the ``deformation parameter" $\beta$ at some typical
angular momentum $L$ (with the parameters in Eq.~(1) being taken
as $\varepsilon_d = 1.010812$~MeV, $A= - 0.153669$~MeV, $B =
0.0822402$~MeV, $C= 0.0185084$~MeV. ). }
\end{center}
\end{figure}

\begin{figure}
\begin{center}
\includegraphics[scale=1.20,angle=0]{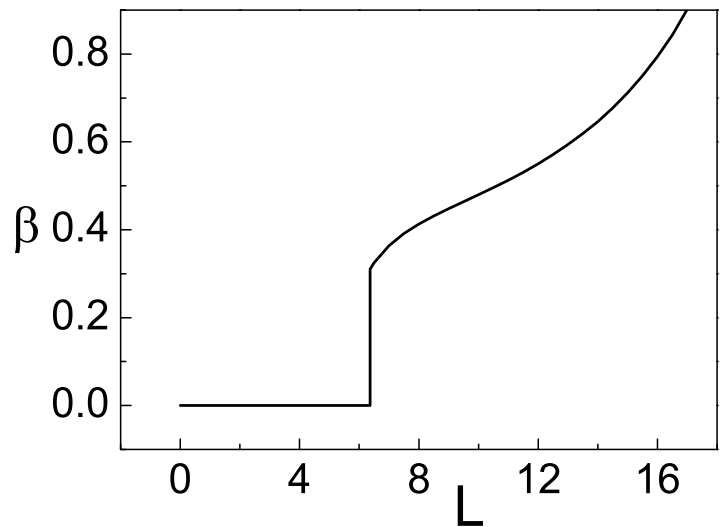}
\caption{The vibration and rotation phase diagram of a nucleus in
terms of the angular momentum $L$ and the ``deformation parameter"
$\beta$ (with the same parameters for Fig.~1). }
\end{center}
\end{figure}

\begin{figure}
\begin{center}
\includegraphics[scale=1.20,angle=0]{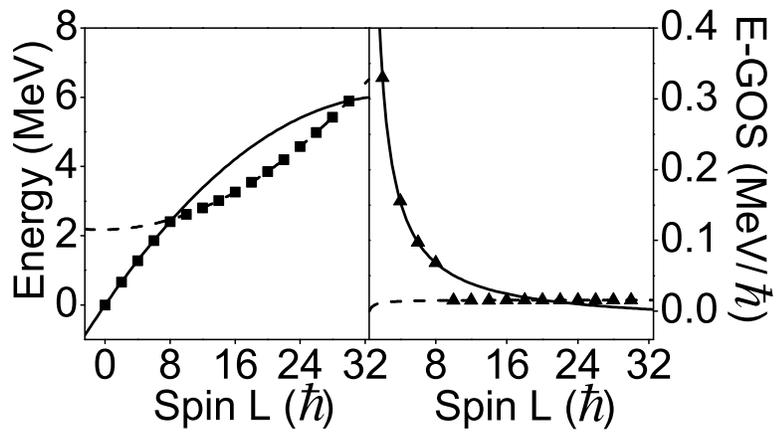}
\caption{An example of the energy against spin (left panel) and
the E-GOS (right panel) along the yrast line (filled squares and
triangles, respectively) in the approach of U(5) symmetry with $(A
+B)< 0$ (with parameters $\varepsilon _d = 0.80$~MeV, $A= -
0.025$~MeV, $B = -0.01$~MeV, $C= 0.004$~MeV. The solid and dashed
lines are implemented to guide the eye.)}
\end{center}
\end{figure}

\vfill

\begin{figure}
\begin{center}
\includegraphics[scale=1.0,angle=0]{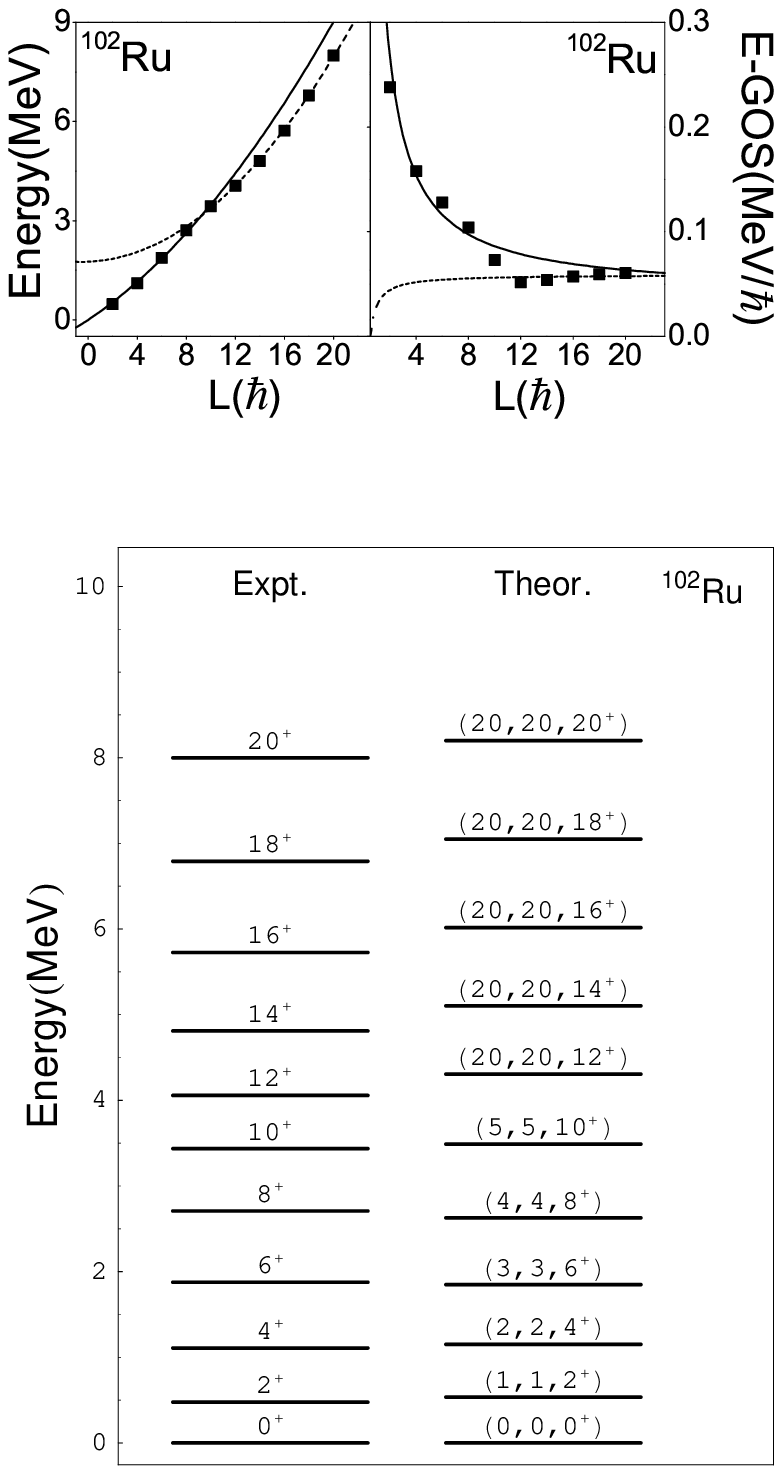}
\caption{The theoretically obtained energy spectrum and E-GOS plot
of the yrast band of $^{102}$Ru(curve) and the comparison with
experimental data (filled squares, taken from Ref.\cite{Regan03})
}
\end{center}
\end{figure}

\begin{figure}
\begin{center}
\includegraphics[scale=1.0,angle=0]{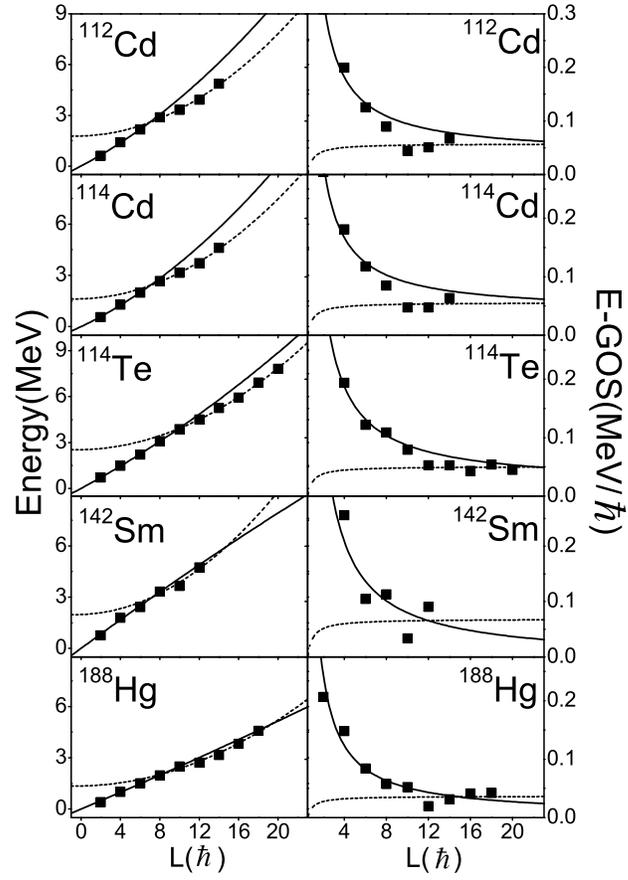}
\caption{The theoretically obtained energy spectrum and E-GOS plot
of the yrast bands of $^{112}$Cd, $^{114}$Cd, $^{114}$Te,
$^{142}$Sm and $^{188}$Hg (curve) and the comparison with
experimental data (filled squares, taken from
Refs.\cite{FJ96,Blac02,Tuli00,Singh02} ) }
\end{center}
\end{figure}

\end{document}